\documentclass[aps,prb,amsmath,amssymb,twocolumn]{revtex4-2}

\usepackage{amsmath, amssymb, bbold}
\usepackage[usenames,dvipsnames]{xcolor}
\usepackage{subfig}
\usepackage{hyperref}
\hypersetup{
	colorlinks,
	citecolor=RoyalBlue,
	filecolor=black,
	linkcolor=ForestGreen,
}
\usepackage{enumitem}
\setlist[itemize]{align=parleft,left=0pt..1em}
\usepackage{comment}

\usepackage{graphicx}
\usepackage{tikz,lipsum,lmodern}

\graphicspath{{./}}

\begin{document}
	
\title{%
Thermodynamics of a Quantum Subsystem
}

\author{Parth Kumar} \email{parthk@arizona.edu}
\author{Charles A.\ Stafford} \email{stafford@physics.arizona.edu}
\affiliation{Department of Physics, University of Arizona, 1118 East Fourth Street, Tucson, Arizona 85721, USA}
\date{\today}
	
\begin{abstract}
Several prior attempts to formulate the Laws of Thermodynamics for a small region within a larger quantum system have led to inconsistencies and unexplained infinities.
The entropy and external work, in particular, require careful analysis when partitioning over the various subsystems.
In this work, we analyze %
the thermodynamics of a quantum subsystem %
driven quasi-statically by external forces.
We show that the thermodynamic functions of a quantum subsystem can be defined dynamically in terms of its local spectrum.
The external work is found to be intrinsically nonlocal due to the nonlocal character of the underlying quantum states. 
This nonlocal quantum work can be harnessed in a ``quantum lever'' to provide up to 100\% amplification of the local work done on a quantum subsystem.

\end{abstract}

\maketitle
\tableofcontents

\section{Introduction}\label{sec_intro}

Thermodynamics in the quantum regime has become a topic of intense investigation in recent years \cite{binderThermodynamicsQuantumRegime2018}. Much of the interest in the topic is driven by the hope of building robust quantum machines with functionalities inaccessible to 
their classical counterparts. The assessment of claims of such ``quantum advantage" (due to the quantum superposition principle and entanglement) being achieved by these machines in the real world requires a rigorous thermodynamic analysis of their performance.

A long-standing debate in the field has been how to formulate the laws of thermodynamics for a quantum subsystem strongly coupled to its environment. A thermodynamic description of the entire universe is well posed \cite{landaulifshitzstatmechbook} %
and (mostly) uncontroversial. 
However, one is typically concerned with scenarios where the given quantum subsystem is described in detail while only a coarse-grained description of environment is specified. Failure to develop the thermodynamics for such a scenario severely limits the whole program of quantum thermodynamics.

Here the issue of partitioning the non-negligible interface arises \cite{talknerColloquiumStatisticalMechanics2020, ludovicoDynamicalEnergyTransfer2014,bruchQuantumThermodynamicsDriven2016,espositoEntropyProductionCorrelation2010,espositoNatureHeatStrongly2015,strasbergFirstSecondLaw2021a,lacerdaQuantumThermodynamicsFast2023,bergmannGreenfunctionPerspective2021,ochoaEnergyDistributionLocal2016,seifertFirstSecondLaw2016,bruchLandauerButtikerApproachStrongly2018, whitneyNonMarkovianQuantumThermodynamics2018} %
and can lead to stark violations and contradictions of thermodynamic principles if not addressed carefully \cite{espositoNatureHeatStrongly2015,talknerColloquiumStatisticalMechanics2020}. Much of the framework of thermodynamics is built on the neglect of the interface, which is usually well-justified for macroscopic systems and is an underlying assumption in the development of textbook thermodynamics \cite{reifbookch2pp9495}. For small systems, be they classical or quantum, this assumption is clearly not valid.

Recently, new light was shed on this debate, wherein it was shown \cite{webbHowPartitionQuantum2024a,kumarWorkSumRule2024a} that the unique division between system and environment leading to a nonsingular subsystem entropy is that based on a partition of Hilbert space. This analysis revealed an unanticipated corollary, namely the nonlocal character of quantum work \cite{kumarWorkSumRule2024a}.
A key focus of the present article is whether the nonlocality of quantum work can be exploited as a new class of quantum advantage.

Whereas Ref.\ \cite{kumarWorkSumRule2024a} treated a single finite quantum system coupled strongly to an infinite environment, %
such as a quantum dot coupled to external macroscopic electrode(s), the present article considers two finite quantum systems strongly coupled to each other, and only weakly coupled to their common environment. For example, a molecular sidegroup covalently bonded to a larger molecule in gas phase.
It is interesting to explore the partitioning of thermodynamic quantities such as work in such driven systems, whose spectra and other properties are distinctly different from those considered in Ref.\ \cite{kumarWorkSumRule2024a}.
 
In this article, we investigate the %
thermodynamic partitioning of finite quantum systems. %
We consider a quasi-statically driven, statistically open but quantum mechanically closed system in the grand canonical ensemble. More specifically, we consider a bipartite quantum system of non-interacting fermions driven quasi-statically by external forces while remaining in weak contact with a reservoir kept at fixed temperature and chemical potential, with which the system may exchange energy and particles. %
We show that the thermodynamics of any subsystem within this larger quantum system \footnote{The thermodynamics of a quantum subsystem in the absence of external work was analyzed in \cite{ptaszynskiEntropyProductionOpen2019}. A slightly generalized analysis was given in \cite{seshadriEntropyInformationFlow2021}.}
can be constructed from the local spectrum of the subsystem or equivalently from the local probability-weighting of global thermodynamic quantities.  

In doing so, the quantum nonlocal character of the external work done is made manifest. This nonlocality of thermodynamic work \cite{kumarWorkSumRule2024a}, inherited from the nonlocality of the underlying quantum states, is encapsulated in a Sum Rule relating the total external work to the Hilbert-space partitioned work done on the subsystems. The phenomenon of nonlocal work in driven quantum systems may be akin to the related but distinct notions of measurement-induced energy teleportation \cite{hottaProtocolQuantumEnergy2008,ikedaDemonstrationQuantumEnergy2023} 
and conditional work at a distance
\cite{elouardInteractionFreeQuantumMeasurementDriven2020} in autonomous quantum systems.

We then illustrate how nonlocal work may be gainfully utilized for ``work amplification" by means of a one-shot quantum lever wherein the external force acts locally on a subsystem yet performs negative work-at-a-distance on the complementary subsystem. It is shown  analytically that such local work amplification can be as much as 100\% under the right parameter tuning and driving protocol.
Finally, we simulate all the partitioned thermodynamic quantities for such levers in two- and four-level quantum systems in contact with a reservoir.   

This paper is organized as follows: In Sec.\ \ref{sec_driven_qsys} we describe the closed quantum system considered in this work and define all the global thermodynamic quantities associated with it. Sec.\ \ref{sec_hilbert_sp_part} defines the subsystem Hilbert-space partitioned thermodynamic quantities, connects them to the local spectrum of the subsystem, and derives the First Law at the level of subsystems. Sec.\ \ref{sec_wsr} elucidates the nonlocal character of the partitioned work by means of the Work Sum Rule. In Sec.\ \ref{sec_lever} we present the analysis of the quantum lever, proving upper bounds on its mechanical advantage and present numerical simulations for the two- and four-level system based levers. Finally, we present our conclusions in Sec.\ \ref{sec_conclusions}.

\section{Closed Quantum System in the Grand Canonical Ensemble}\label{sec_driven_qsys}

\begin{figure}
\centering
\begin{tikzpicture}[scale=1]%

            \draw[fill=green!20] (5.5,1.5) circle (1cm);
			\node at (5.5,1.5) {$H_{\bar{S}}$};

			\draw[fill=green!20] (2,1.5) circle (1cm);
			\node at (2,1.5) {$H_{S}(t)$};
			\draw[<->](3,1.5)--(4.5,1.5);
			\node[below] at (3.7,1.5) { $H_{S\bar{S}}(t)$};

   \draw[thick,fill=red!20] (1,-1)--(1,0)-|(6.5,-1);
   \draw[<->,dashed](2,0)--(2,0.5);
   \draw[<->,dashed](5.5,0)--(5.5,0.5);

   \node at (3.7,-0.5) {$T\,,\mu$};

		\end{tikzpicture}
  \captionsetup{justification=raggedright, singlelinecheck=false}
  \caption{Schematic for a finite quantum system, consisting of subsystems $S$ and $\bar{S}$, in the Grand Canonical Ensemble [Eq.\ \eqref{eq_closedsys_ham_gen}]. %
  }
\label{fig_cqs_gce_schm}
\end{figure}
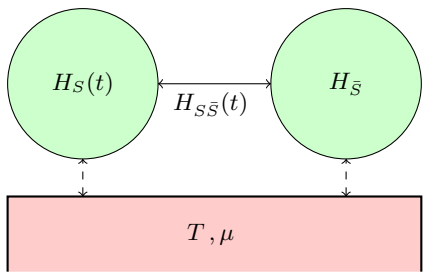

In this work, we consider a quasi-statically driven, quantum mechanically closed but statistically open system of independent fermions weakly coupled to a single fermionic reservoir with which it exchanges particles and energy (see Fig.\ \ref{fig_cqs_gce_schm} for a schematic). The reservoir is maintained in equilibrium at constant temperature $T$ and chemical potential $\mu$ throughout the drive. For such a setup, quasi-static driving means that the Hamiltonian parameter(s) are driven slowly enough that the eigenstates of the closed quantum system are populated, at all times, according to the Fermi-Dirac distribution of the reservoir (see Appendix \ref{app_full_ham_qstat_drive} for details). %
The Fock-space Hamiltonian of such a system can be written as
\begin{equation}\label{eq_closedsys_ham_gen}
    H(t)=H_S(t)+H_{\bar{S}}+H_{S\bar{S}}(t)\,,
\end{equation} 
where %
\begin{equation}
   H_S(t)=\sum_{n,m}[h_{S}(t)]_{nm}d^{\dagger}_{n}d_{m}\,,
\end{equation}
\begin{equation}
   H_{\bar{S}}=\sum_{i,j}[h_{\bar{S}}]_{ij}c^{\dagger}_{i}c_{j}\,,
\end{equation}
and 
\begin{equation}
   H_{S\bar{S}}(t)=\sum_{i,n}\Bigg([h_{S\bar{S}}(t)]_{in}c_{i}^{\dagger}d_{n}+\mbox{h.c.}\Bigg)\,.
\end{equation}
Here $[h_S(t)]_{nm}$, $[h_{\bar{S}}]_{ij}$, and $[h_{S\bar{S}}(t)]_{in}$ are matrix elements of Hamiltonians acting on the single-particle Hilbert-space of the subsystems $S$, $\bar{S}$, and the coupling between them,  respectively, where we have allowed for explicit time-dependence in the subsystem $S$ and coupling Hamiltoninans. The operators $d^\dagger_n$ ($d_n$) and $c^\dagger_i$ ($c_i$) are fermionic creation (annihilation) Fock-space operators satisfying $\{d_n,d^\dagger_m\}=\delta_{nm}$,
$\{c_i,c^\dagger_j\}=\delta_{ij}$, and $\{d_n,d_m\}=\{c_i,c_j\}=\{d_n,c_i\}=0$. Although the coupling of the composite quantum system to the reservoir is weak, there is no a priori reason to expect weak coupling between the quantum subsystems, i.e. subsystems $S$ and $\bar{S}$ can be strongly coupled via $H_{S\bar{S}}$, or in other words energy scales of $H_{S\bar{S}}$ may be comparable to those of $H_S$ and $H_{\bar{S}}$. 

Throughout this paper we use the following notation for quantum operators: Any general one-body operator in Fock-space is denoted as
\begin{equation}
    O(t)=\sum_{k,l}o_{kl}(t)a^{\dagger}_{k}a_l +\mbox{h.c.} \,,
\end{equation}
where $o_{kl}(t)=\langle k|o(t)|l\rangle$ is the matrix element of the observable $o(t)$ defined on the single-particle Hilbert space and the Fock-space creation operator $a^{\dagger}_{i} \in \{c^{\dagger}_{i},d^{\dagger}_{i}\ \ \forall\ \ i\}$. The uppercase $O(t)$ thus denotes the Fock-space operator corresponding to the lowercase $o(t)$ single-particle Hilbert space operator.

The Internal Energy $U(t)$ of a quantum system with Hamiltonian $H(t)$ is %
\begin{equation}\label{eqn_integy_gen_def}
    U(t):=\langle H(t) \rangle\,,
\end{equation}
where $\langle\,\rangle$ denotes the quantum statistical average $\langle H(t) \rangle=\mathrm{Tr}\{H(t)\rho(t)\}$, where $\rho(t)$ is the density matrix of the universe at time $t$ and $\mathrm{Tr}\{\}$ denotes the trace over the full Fock space. The (inclusive) rate of Work done by external forces on the quantum system is  \cite{jarzynskiComparisonFarfromequilibriumWork2007,campisiColloquiumQuantumFluctuation2011,talknerColloquiumStatisticalMechanics2020}
\begin{equation}\label{eqn_totalextpower_def}
  \dot{W}_{ext}(t):=%
  \langle \dot{H}(t) \rangle\,.
\end{equation}
Finally, we have the von Neumann Entropy of the system 
\begin{equation}
    S(t):=\langle \mathcal{S}(t) \rangle= -\mathrm{Tr}\{\rho(t)\ln\rho(t)\}\,,
\end{equation}
where the entropy operator $\mathcal{S}(t)=-\ln\rho(t)$. 

According to the quantum adiabatic theorem \cite{Sakurai2011}, the solution to the Schr\"odinger equation for a system with a %
quasi-statically driven Hamiltonian $h(t)$ is given by a linear combination of vectors of the form 
\begin{equation}
    |\psi_{\nu}(t)\rangle=e^{i\theta_{\nu}(t)}e^{i\gamma_{\nu}(t)}|\nu(t)\rangle\,,
\end{equation}
where $\theta_{\nu}(t)=-\frac{1}{\hbar}\int_{0}^{t}dt'\,\epsilon_{\nu}(t')$ is the so-called dynamical phase,  $\gamma_{\nu}(t)=i\int_{0}^{t}dt'\,\langle\psi_{\nu}(t')|\frac{\partial}{\partial t'}\psi_{\nu}(t')\rangle$ is the so-called geometrical phase, and $|\nu(t)\rangle$ solves the instantaneous Hamiltonian eigenvalue equation $h(t)|\nu(t)\rangle=\epsilon_{\nu}(t)|\nu(t)\rangle$. 
Under quasi-static driving, these instantaneous eigenstates are populated according to the equilibrium Fermi-Dirac distribution of the reservoir, so the 
density matrix of the system is 
\begin{equation}
    \rho^{(0)}(t)= \prod_\nu [f_\nu(t) \psi_\nu^\dagger(t)\psi_\nu(t) + (1-f_\nu(t))\psi_\nu(t)\psi_\nu^\dagger(t)],
\end{equation}
where $\psi_\nu^\dagger(t)$ creates a fermion in an eigenstate of the instantaneous Hamiltonian $h(t)$, $\psi_\nu^\dagger(t)|0\rangle = |\nu(t)\rangle$, and $f(\epsilon)=(1+e^{\beta(\epsilon-\mu)})^{-1}$, where $\beta^{-1}=T$ (throughout this work we set $k_B=1$) is the Fermi-Dirac distribution function and we have written $f_\nu(t) \equiv f(\varepsilon_\nu(t))$ to lighten notation. %

Using the above definitions, the relevant thermodynamic quantities for a quasi-statically driven system can now be evaluated as follows. %
The internal energy is given by
\begin{equation}\label{eqn_qstat_integy_dos_rel}
 U^{(0)}(t)=\sum_{\nu}f_\nu(t)\epsilon_{\nu}(t)\,.
\end{equation} 
Here the superscript $U^{(n)}$ denotes the order in time derivatives of the driving Hamiltonian.

Similarly, the quasi-static power delivered may be computed as %
\begin{subequations}\label{eqn_qstat_extpower_spec_rel}
\begin{eqnarray}\label{eqn_qstat_extpower_spec_rel_a}
    \dot{W}_{ext}^{(1)}(t)&=&\sum_{\nu}f_\nu(t)\langle\nu(t)|\dot{h}(t)|\nu(t)\rangle %
    \\ \label{eqn_qstat_extpower_spec_rel_b}
&=&\sum_{\nu}f_\nu(t)\dot{\epsilon}_\nu(t) \,,    
\end{eqnarray}
\end{subequations}
where Eq.\ \ref{eqn_qstat_extpower_spec_rel_b} follows from the adiabatic theorem, or more generally, from the Hellman-Feynman theorem.
The quasi-static entropy is given by 
\begin{eqnarray}\label{eqn_totalentropy_dos_rel}
 S^{(0)}(t)=-\sum_{\nu} \{f_\nu(t)\ln f_\nu(t) + %
 (1-f_\nu(t))\ln[1-f_\nu(t)]\}  \,. \nonumber\\
\end{eqnarray} 
The mean number of particles in the system is given by
\begin{equation}\label{eqn_totalpartnum_dos_rel}
 N^{(0)}(t)=\sum_{\nu}f_\nu(t)\,,
\end{equation}
The grand canonical potential %
of the system is %
\begin{equation}\label{eqn_grandpotdef}
\Omega^{(0)}(t)=U^{(0)}(t)-TS^{(0)}(t)-\mu N^{(0)}(t) \,,
\end{equation}
which can be expressed as
\begin{equation} \label{eqn_totalgrandpot_dos_rel}
 \Omega^{(0)}(t)= -\frac{1}{\beta}\sum_{\nu}\ln[1+e^{-\beta(\epsilon_{\nu}(t)-\mu)}]\,.
\end{equation}

It can be shown that the first variations of the above thermodynamic quantities satisfy \cite{kumarWorkSumRule2024a} %
\begin{equation}\label{eqn_global_fdmt_thermo_id}
    \dot{\Omega}^{(1)}(t)=\dot{U}^{(1)}(t)-T\dot{S}^{(1)}(t)-\mu \dot{N}^{(1)}(t)\,,
\end{equation}
where 
the time derivative of the grand potential is given by %
\begin{equation}
    \dot{\Omega}^{(1)}(t)=\sum_{\nu}f_\nu(t)\dot{\epsilon}_\nu(t) 
\end{equation}
and the reversible heat transferred from the reservoir is given by
\begin{equation}
    T\dot{S}^{(1)}(t)=\sum_{\nu}(\epsilon_\nu(t)-\mu)\dot{f}_\nu(t)\,.
\end{equation}    
It then also follows from Eq.\ \eqref{eqn_qstat_extpower_spec_rel} that the first variation of $\Omega$ is equal to the external work \cite{kumarWorkSumRule2024a} %
\begin{equation}\label{eqn_omega_work_global_rel}
    \dot{W}_{ext}^{(1)}(t)=\dot{\Omega}^{(1)}(t).
\end{equation}

\section{Subspace Thermodynamic Quantities}\label{sec_hilbert_sp_part}

We define the thermodynamic quantities of subsystem $\gamma$ as the quantum statistical averages of partitioned quantum observables $H|_\gamma$, $S|_\gamma$, and $N|_\gamma$, respectively, where %
$O|_\gamma$ is the Fock-space operator corresponding to the following operator defined on the single-particle Hilbert-space \cite{webbHowPartitionQuantum2024a,kumarWorkSumRule2024a} %
\begin{equation}\label{eqn_hilb_sp_partition_def}
o|_{\gamma}=\frac{1}{2}\{\pi_{\gamma},o\}\,,
\end{equation} 
where $\pi_{\gamma}=\int_{x\in\gamma}\,dx |x\rangle\langle x|$ is the projection operator onto subspace $\gamma$ of the single-particle Hilbert-space,
and $o$ is the single-particle Hilbert-space operator corresponding to the global Fock-space operator $O=\sum_\gamma O|_\gamma$, while the anticommutator 
(defined as $\{a,b\}=ab+ba$) ensures the hermiticity of $O|_\gamma$.
For $\gamma=S$, %
definition \eqref{eqn_hilb_sp_partition_def} implies
\begin{equation}
    H|_S(t)=H_{S}(t)+\frac{1}{2}H_{S\bar{S}}(t),
\end{equation}
so that the coupling Hamiltonian is partitioned equally between the two subsystems. 

The local thermodynamic quantities so constructed are given  %
by weighting the contribution of each eigenstate to a given global quantity [Eqs.\ \eqref{eqn_qstat_integy_dos_rel}, \eqref{eqn_totalentropy_dos_rel}, \eqref{eqn_totalpartnum_dos_rel}, and \eqref{eqn_totalgrandpot_dos_rel}] by the probability  of finding a particle in the $\nu^{th}$ eigenstate within the subspace $\gamma$ (see Appendix \ref{app_hilbertspace_part_derv} for details)
\begin{equation}\label{eqn_prob_nu_gamma_def}
     P_{\nu}(\gamma,t):= \langle\nu(t)|\pi_\gamma|\nu(t)\rangle=\int_{x\in\gamma}\,dx\,|\psi_{\nu}(x,t)|^2 \,,
\end{equation}
with $\psi_{\nu}(x,t)$ an instantaneous eigenfunction of $h(t)$. %

The local thermodynamic quantities of subsystem $\gamma$ are
\cite{staffordLocalEntropyNonequilibrium2017,shastryThirdLawThermodynamics2019,shastryTheoryThermodynamicMeasurements2019} %
\begin{equation}\label{eqn_part_energy_def}
 U_{\gamma}^{(0)}(t)=\sum_{\nu}P_{\nu}(\gamma,t)f(\epsilon_{\nu}(t))\epsilon_{\nu}(t)\,,
\end{equation}
\begin{equation}\label{eqn_part_entropy_def}
 S_{\gamma}^{(0)}(t)=\sum_{\nu}P_{\nu}(\gamma,t)s(\epsilon_{\nu}(t))\,,
\end{equation}
where $s(\epsilon)=\beta(\epsilon-\mu)f(\epsilon)+\ln(1+e^{-\beta(\epsilon-\mu)})$, 
\begin{equation}\label{eqn_part_partnum_def}
N_{\gamma}^{(0)}(t)=\sum_{\nu}P_{\nu}(\gamma,t)f(\epsilon_{\nu}(t))\,,
\end{equation}
and 
\begin{equation}\label{eqn_part_grandpot_def}
 \Omega_{\gamma}^{(0)}(t)=\sum_{\nu}P_{\nu}(\gamma,t)\omega(\epsilon_{\nu}(t))
\end{equation}
where $\omega(\epsilon)=-\frac{1}{\beta}\ln(1+e^{-\beta(\epsilon-\mu)})$.
It follows straightforwardly from the above definitions of subspace quantities that the global thermodynamic identity [Eq.\ \eqref{eqn_global_fdmt_thermo_id}] holds at the level of subsystems
\begin{equation}\label{eqn_thermo_id_part}
 \dot{\Omega}_{\gamma}^{(1)}(t)=\dot{U}_{\gamma}^{(1)}(t)-T\dot{S}_{\gamma}^{(1)}(t)-\mu \dot{N}_{\gamma}^{(1)}(t)\,,   
\end{equation}
where a summation over $\gamma$ of the above identity recovers the global identity [Eq.\ \eqref{eqn_global_fdmt_thermo_id}].

Rearranging the terms
in Eq.\ \eqref{eqn_thermo_id_part}, and %
defining
$\dot{W}_\gamma^{(1)}(t):=\dot{\Omega}_{\gamma}^{(1)}(t)$ as the {\it rate of thermodynamic work} on subsystem $\gamma$, we
obtain the First Law of Thermodynamics for subsystem $\gamma$
\begin{equation}\label{eqn_1st_Law_gamma}
    \dot{U}_{\gamma}^{(1)}(t)=T\dot{S}_{\gamma}^{(1)}(t)+\mu \dot{N}_{\gamma}^{(1)}(t)+\dot{W}_\gamma^{(1)}(t)\,.
\end{equation}

\section{Work Sum Rule
}\label{sec_wsr}

The central result of this work is that Eq.\ \eqref{eqn_omega_work_global_rel} can be rewritten using the Hilbert-space partition introduced in the previous section as 
\begin{equation}\label{eqn_closedsys_worksumrule}
    \dot{W}_{ext}^{(1)}(t)=\dot{\Omega}_S^{(1)}(t)+\dot{\Omega}_{\bar{S}}^{(1)}(t) \,.
\end{equation}
We term the above result the {\it sum rule} for quantum work, analogous to that for open quantum systems \cite{kumarWorkSumRule2024a}. 
Separately, it follows from Eq. \eqref{eqn_totalextpower_def} that %
\begin{equation}
  \dot{W}_{ext}^{(1)}(t)=\langle \dot{H}|_S(t)\rangle + \langle\dot{H}|_{\bar{S}}(t)\rangle  \,.
\end{equation}
However, it can be seen from explicit evaluations that the expectation value of the power operator partitioned on that subsystem is not, in general, equal to the rate of thermodynamic work done on a given subsystem 
\begin{equation}
  \langle\dot{H}|_\gamma\rangle \neq \dot{\Omega}_{\gamma}^{(1)}(t)\,.
\end{equation}
Instead, %
\begin{equation}\label{eqn_omegacurr_def}
\dot{W}^{(1)}_{\gamma}(t) :=\dot{\Omega}_{\gamma}^{(1)}(t)
=\langle \dot{H}|_{\gamma}(t)\rangle+I^W_\gamma(t)\,, 
\end{equation}
where the difference
$I^W_{\gamma}(t)$ represents the rate of nonlocal quantum work. 

Each term appearing in Eq.\ \eqref{eqn_omegacurr_def} can be evaluated explicitly. For the rate of thermodynamic work done on subsystem $\gamma$ we insert Eq.\ \eqref{eqn_prob_nu_gamma_def} into Eq.\ \eqref{eqn_part_grandpot_def}, and take the time derivative, obtaining 
\begin{eqnarray}\label{eqn_part_grandpot_prob_rel}
\dot{W}^{(1)}_\gamma(t)&:=&\dot{\Omega}_{\gamma}^{(1)}(t)=\sum_{\nu}[\dot{P}_{\nu}(\gamma,t)\omega_{\nu}+P_{\nu}(\gamma,t)\dot{\omega}_{\nu}] \nonumber \\ &=&\sum_{\nu}[\dot{P}_{\nu}(\gamma,t)\omega_{\nu}+P_{\nu}(\gamma,t)f_\nu\dot{\epsilon}_{\nu} ]\,,
\end{eqnarray}
where $\omega_\nu=\omega(\epsilon_\nu)$.
The expectation value of the partitioned power can be computed as (see Appendix \ref{app_closedsys_extpower_prob_rel_derv} for details)
\begin{eqnarray}\label{eqn_part_extwork_prob_rel}
\langle \dot{H}|_\gamma(t)\rangle &=&\sum_{\nu}P_{\nu}(\gamma,t)f_\nu\dot{\epsilon}_{\nu}\nonumber \\ &+& \sum_{\mu\neq\nu}\frac{f_\nu+f_\mu}{2}\langle\nu|\dot{h}|\mu\rangle\langle\mu|\pi_{\gamma}|\nu\rangle \,. 
\end{eqnarray}
Thus the rate of nonlocal work %
$I^W_{\gamma}(t)$ implicity defined in Eq.\ \eqref{eqn_omegacurr_def} is %
\begin{eqnarray}\label{eqn_part_omegacurr_prob_rel}
 I^W_{\gamma}(t)&=&\sum_{\nu}\dot{P}_{\nu}(\gamma,t)\omega_{\nu} \nonumber \\ && -
 \sum_{\mu\neq\nu}\frac{f_\nu+f_\mu}{2}\langle\nu|\dot{h}|\mu\rangle\langle\mu|\pi_{\gamma}|\nu\rangle\,.
\end{eqnarray}
This term is subject to a conservation condition over the subsystems: $\sum_{\gamma}I^W_{\gamma}=0$. %
$I^W_{\gamma}$ may be thought of as the \emph{instantaneous rate of quantum work at a distance} on subsystem $\gamma$.  The first term in Eq.\ \eqref{eqn_part_omegacurr_prob_rel} represents the flow of free energy between subsystems, while the second term can be interpreted as the rate at which energy is generated nonlocally in the subsystem. %

The nonlocal character of quantum work stems from the nonlocality of the states of the quantum system.  $I^W_{\gamma}(t)$ 
is due to the time evolution of the quasi-stationary states of the system, and is to be contrasted with the usual energy transfer mediated by the coupling $H_{S\bar{S}}$ (see Appendix \ref{app_nonlocalwork_vs_usualegytransfer} for details).
The role of extended states is obvious in the first term on the RHS of Eq.\ \eqref{eqn_part_omegacurr_prob_rel}, as the probability distributions of the instantaneous energy eigenstates evolve with the changing Hamiltonian.  The second term on the RHS of Eq.\ \eqref{eqn_part_omegacurr_prob_rel} describing nonlocal energy generation is more subtle to interpret.  It is generically nonzero because the instantaneous energy eigenstates, while mutually orthogonal on the entire Hilbert space, have a nonzero overlap on the subsystem Hilbert space 
($\langle \mu|\pi_\gamma|\nu\rangle\neq0$).  However, this term, like the first term on the RHS of Eq.\ \eqref{eqn_part_omegacurr_prob_rel}, sums to zero over the subsystems.

The nonlocal character of quantum work is most clearly manifest when
only the subsystem Hamiltonian is driven ($\dot{H}=\dot{H}_S$); the instantaneous rate of quantum work on the complementary subsystem is then entirely due to nonlocal work
\begin{equation}\label{eqn_sum_rule_local}
I^W_{\bar{S}}(t)
\stackrel{\dot{H}_{S\bar{S}}=0}{=} \dot{W}_{\bar{S}}(t)=
   \langle \dot{H}_S(t)\rangle - \dot{\Omega}_S^{(1)}(t) \,,
\end{equation}
and is generically nonzero.

\section{Results for Model driven quantum systems}\label{sec_lever}

\subsection{Time-dependent two-level system}\label{tls_subsec}

\begin{figure}
    \centering
    \includegraphics[width=0.45\textwidth,height=19cm]{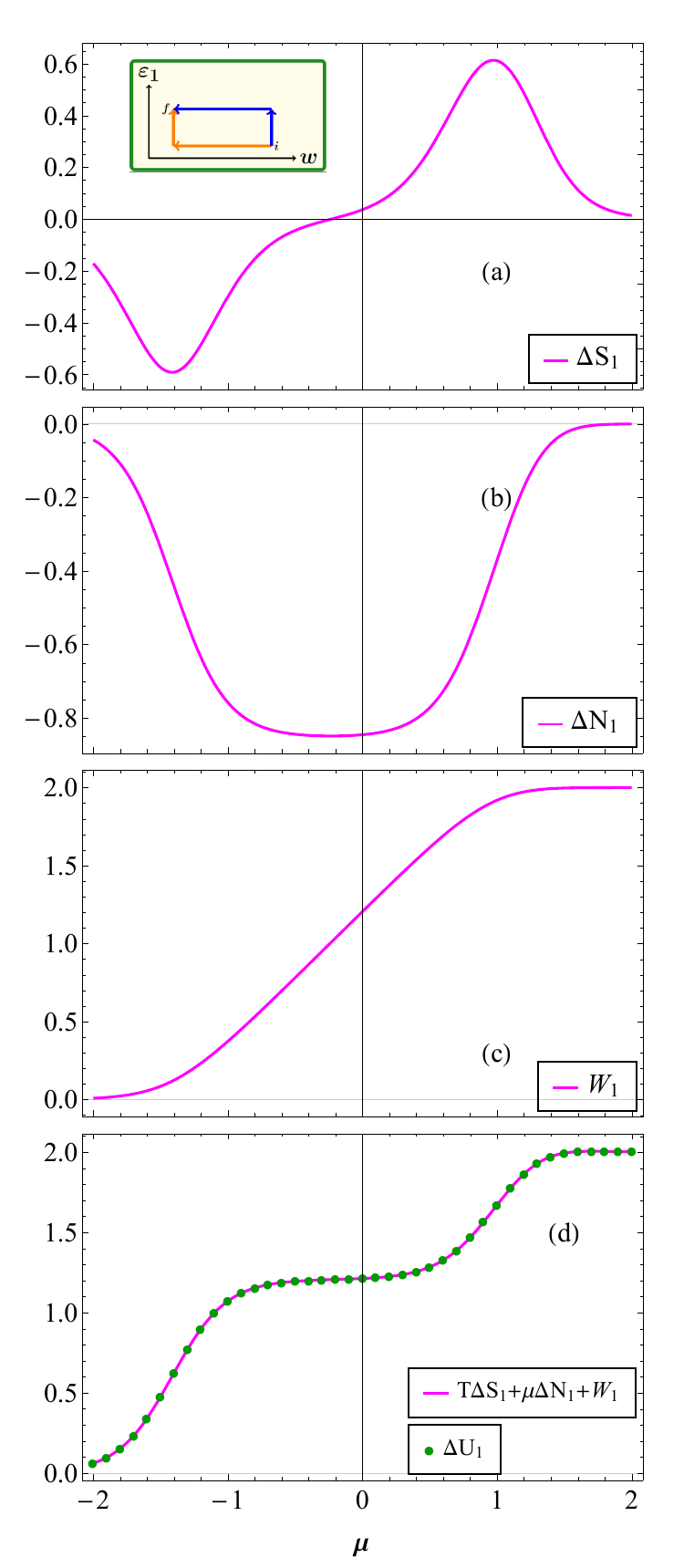}
    \captionsetup{justification=raggedright, singlelinecheck=false}
    \caption{The integrated terms (a) entropy $\Delta S_1$, (b) occupancy $\Delta N_1$, and (c) work $W_1$
in the two-level model under driving protocol 2, as functions of the chemical potential of the reservoir. (d) Verification of the equality of the LHS and RHS of Eq.\ \eqref{eqn_1st_Law_gamma}.  Here the reservoir temperature $T=0.2$, while the Hamiltonian parameters are varied along the two paths shown in the inset in panel (a) with $\varepsilon_1$: $-1\rightarrow 1$, $w$: $1\rightarrow 0.1$, while keeping the other site energy fixed at $\varepsilon_2=1$ .}
\label{fig_tls_E1Vdriven_firstlaw_grid}
\end{figure} 

\begin{figure}
\centering
\includegraphics[width=.45\textwidth,height=15cm]{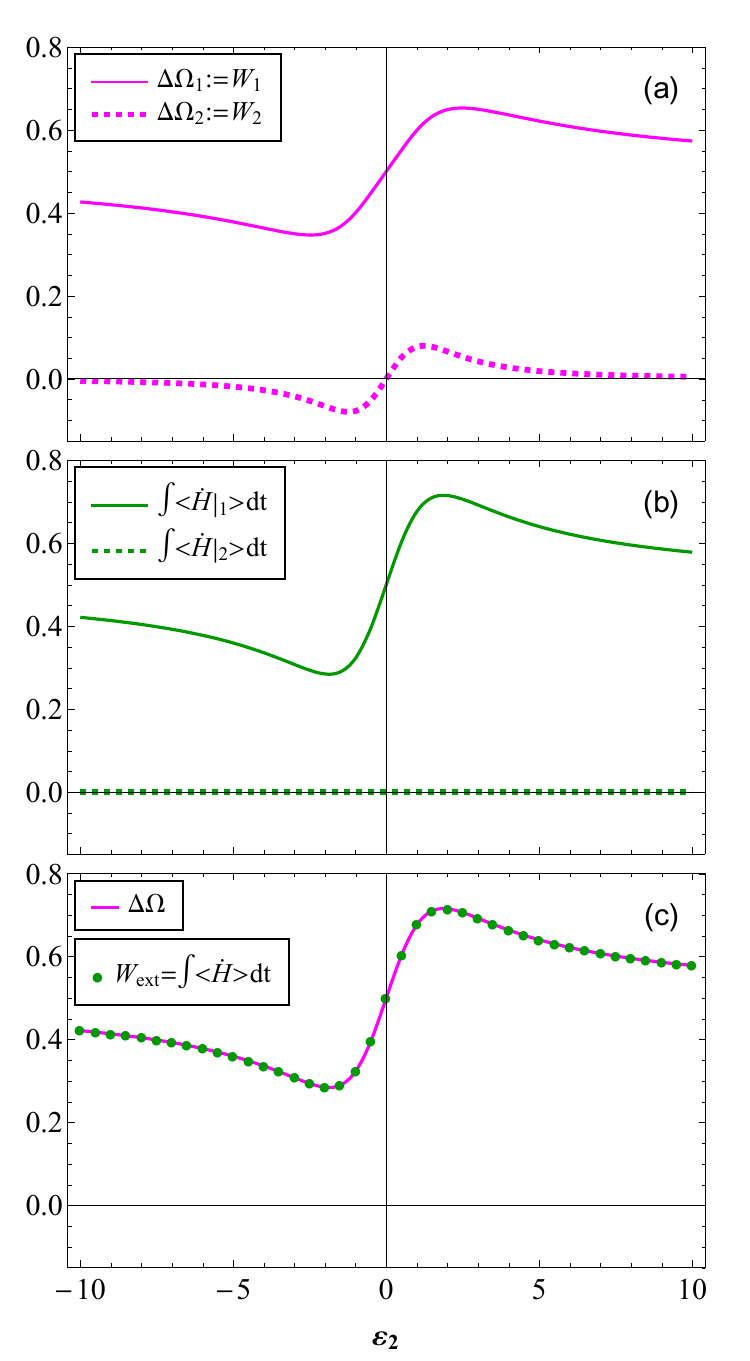}
\captionsetup{justification=raggedright, singlelinecheck=false}	
\caption{Verification of the Work Sum Rule [Eqs.\ \eqref{eqn_closedsys_worksumrule}, \eqref{eqn_sum_rule_local}] for the two-level system in the grand canonical ensemble ($T=0.2$, $\mu=0$), under driving protocol 1.
Only $\varepsilon_1(t)$ is driven (from -0.5 to 0.5) and $w=1$. (a) Changes of the partitioned grand potential, (b) integrals of the partitioned power operator, and (c) comparison of the total external work and change in the global grand potential.
}
\label{fig_tls_workgrandpot_e1driven}
\end{figure}

\begin{figure}%
	
\centering
 \includegraphics[width=.45\textwidth,height=15cm]{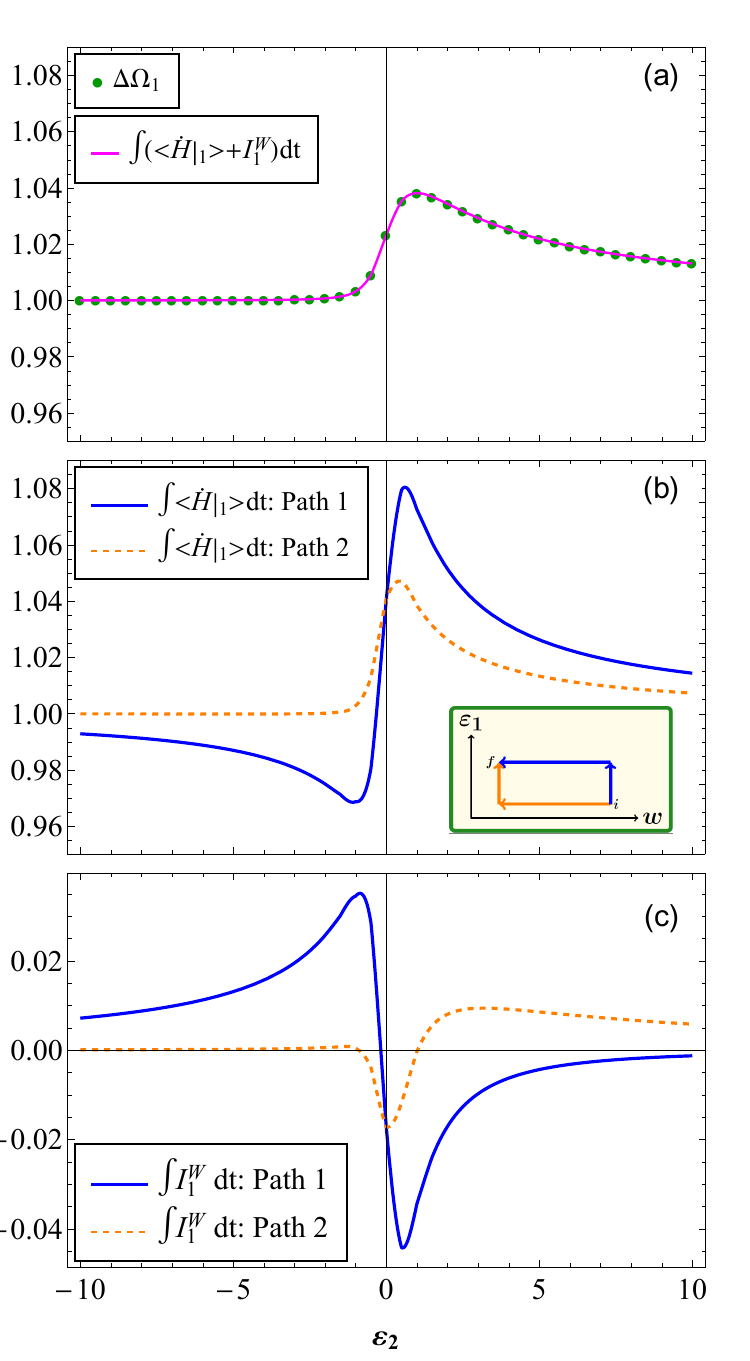}%
	
\captionsetup{justification=raggedright, singlelinecheck=false}	
\caption{Analysis of nonlocal quantum work in a two-level system in the grand canonical ensemble ($T=0.2$, $\mu=0$), driven under protocol 2.
Both the level $\varepsilon_{1}(t)$ and the coupling $w(t)$ are driven along two different paths (see inset: %
$\varepsilon_1$: $-1\rightarrow 1$, $w$: $0.4\rightarrow 0.04$).
(a) Comparison of the LHS and RHS of Eq.\ \eqref{eqn_omegacurr_def}.
(b) Integral of the partitioned power operator on site 1, %
computed from Eq.\ \eqref{eqn_part_extwork_prob_rel}. (c) Nonlocal quantum work on site 1, %
computed from Eq.\ \eqref{eqn_part_omegacurr_prob_rel}.
}
\label{fig_tls_workgrandpot_e1wdriven}
\end{figure}

The Hamiltonian for this model %
is given by Eq.\ \eqref{eq_closedsys_ham_gen} with
\begin{equation}
 H_{S}(t)=\varepsilon_{1}(t)d^{\dagger}d \,, 
 \end{equation}
\begin{equation} 
 H_{\bar{S}}=\varepsilon_{2}c^{\dagger}c \,,
\end{equation}
and 
\begin{equation}
H_{S\bar{S}}(t)=w(t)d^{\dagger}c + \mbox{h.c.} \,. 
\end{equation}
The system is driven slowly enough that it remains in equilibrium with the (weakly coupled) reservoir at temperature $T$ and chemical potential $\mu$ throughout the driving protocol. 
We consider two %
protocols, which consist of quasi-static variations of the parameters of the two-level system: In {\it protocol 1}, only the level 
$\varepsilon_1(t)$ is driven, while in {\it protocol 2}, both the level $\varepsilon_1(t)$ and the coupling $w(t)$ are driven. %
For all the results presented in this subsection, we plot the time-integrals of the partitioned thermodynamic rates %
that appear in the First Law, Eq.\ \ref{eqn_1st_Law_gamma}. Thus we have for the subsystem 1: the change in entropy
$\Delta S_1=\int_{-\infty}^{\infty}\dot{S}_1^{(1)}(t)\,dt$, the change in particle number $\Delta N_1=\int_{-\infty}^{\infty}\dot{N}_1^{(1)}(t)\,dt$, the thermodynamic work done $W_1=\int_{-\infty}^{\infty}\dot{\Omega}_1^{(1)}(t)\,dt$, and the change in internal energy $\Delta U_1=\int_{-\infty}^{\infty}\dot{U}_1^{(1)}(t)\,dt$. The total external work done is denoted by  $W_{ext}=\int_{-\infty}^{\infty}\langle\dot{H}(t)\rangle\,dt = \Delta \Omega$. (See Appendix \ref{app_tls_exact_exprs} for how to compute closed expressions for these quantities.)

The partitioned thermodynamic quantities for subsystem $S$ are plotted as a function of the chemical potential $\mu$ of the reservoir for driving protocol 2 in Fig.\ \ref{fig_tls_E1Vdriven_firstlaw_grid} (we also present the analogous results for protocol 1 in Appendix \ref{app_part_thermo_prot1}).  
All the thermodynamic quantities are path independent under this protocol carried out at constant values of the reservoir chemical potential $\mu$ and temperature $T$.  The First Law of Thermodynamics is verified in Fig.\ \ref{fig_tls_E1Vdriven_firstlaw_grid}(d) by comparison of the integrated right- and left-hand sides of Eq.\ \eqref{eqn_1st_Law_gamma}. 
The resonances and step-like features in the partitioned thermodynamic quantities %
can be understood in terms of the bonding and antibonding orbitals of the 2-level system crossing $\mu$ at the beginning or end of the protocol (see Appendix \ref{app_tls_thermo_fig_discuss} for additional discussions).

The non-local nature of quantum work is most clearly manifested in protocol 1, where the power operator has no off-diagonal contribution.
Figures \ref{fig_tls_workgrandpot_e1driven}(a) and \ref{fig_tls_workgrandpot_e1driven}(b) show the changes in the grand canonical potential partitioned on  subsystems 1 and 2, and the
integrated expectation values of the partitioned power operator, respectively, under protocol 1, wherein $\varepsilon_1$ is varied from -0.5 to 0.5. The net changes of all quantities are plotted as functions of the fixed level $\varepsilon_2$, showing the effects of hybridization when $\varepsilon_2 \approx \varepsilon_1$. 
While the power operator [Fig.\ \ref{fig_tls_workgrandpot_e1driven}(b)] has no contribution in subsystem 2, the instantaneous rate of quantum work on subsystem 2 is clearly nonzero [Fig.\ \ref{fig_tls_workgrandpot_e1driven}(a)].  Adding up the values for the two subsystems, the
Work Sum rule, Eq.\ \eqref{eqn_closedsys_worksumrule}, is verified in Fig.\ \ref{fig_tls_workgrandpot_e1driven}(c).

The changes of the relevant quantities of subsystem 1 under protocol 2, and their path dependence or lack thereof, are illustrated in
Fig.\ \ref{fig_tls_workgrandpot_e1wdriven}.  $\Delta \Omega_1=W_1$ is independent of which path is taken, as is the sum $\int(\langle \dot{H}|_1\rangle + I^W_1)dt$,
as shown in Fig.\ \ref{fig_tls_workgrandpot_e1wdriven}(a).  However, the individual terms $\int\langle \dot{H}|_1\rangle dt$ and $\int I^W_1 dt$ are path
dependent, as shown in Figs.\ \ref{fig_tls_workgrandpot_e1wdriven}(b) and (c), respectively.  The rate of nonlocal quantum work $I^W_1$ is appreciable when there is substantial hybridization of the two levels.

\subsection{Quantum Lever analysis}

The phenomenon of nonlocal quantum work can be leveraged to produce work amplification in a subspace when the nonlocal work on the complementary subspace is negative.  Consider a local external force acting on a quantum subspace; if the nonlocal work generated by the force outside the subspace is negative, then the work on the subspace of interest is greater than the total external work, by conservation of the total energy.  This effect may be interpreted as a ``quantum lever.''  However, unlike a classical lever, which amplifies force but conserves work, the quantum lever amplifies both force and work (on the subspace).  This effect is not a violation of the First Law since cost of local work amplification is being paid elsewhere as negative work in the complementary subspace--total energy is very much conserved.  Moreover, the quantum lever is a single shot device; no such work amplification is possible for a cyclic process.
We define the instantaneous mechanical advantage of such a such a quantum lever as
\begin{equation}\label{eqn_mech_adv_def}
  \eta:=\dot{W}_{1}/\dot{W}_{\text{ext}}\,.  
\end{equation}

\begin{figure}
    \centering
    \includegraphics[width=9cm,height=6.5cm]{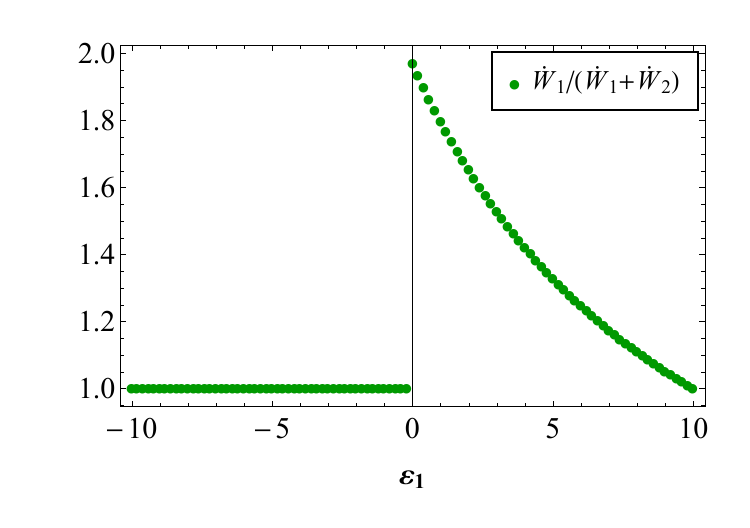}
    \captionsetup{justification=raggedright, singlelinecheck=false}
    \caption{Instantaneous mechanical advantage for the two-level quantum lever where only the level $\varepsilon_1$ is driven (protocol 1). Here the temperature $T=0.0001$, chemical potential $\mu=0$, coupling $w=1$, and $\varepsilon_2=-10$. }
    \label{fig_tls_e1driven_mechadv}
\end{figure}

Figure \ref{fig_tls_e1driven_mechadv} shows the mechanical advantage of a single-shot quantum lever based on a two-level system, where level 1 is driven and the work on subspace 1 is computed.  As is evident from the figure, a maximum mechanical advantage $\eta\approx2$ is realizable for this system.
Indeed, one can show based on a perturbative analysis in the low-temperature limit that for the local driving protocol considered $\eta\leq2$, or in other words, we can get as much as a $100\%$ work amplification (see Appendix \ref{app_lever_bound} for details). 

\begin{figure}
    \centering
    \includegraphics[width=9cm,height=6.5cm]{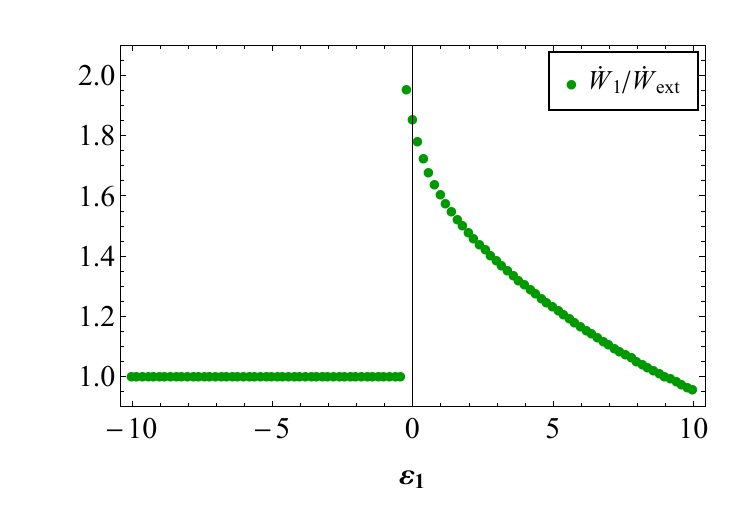}
    \captionsetup{justification=raggedright, singlelinecheck=false}
    \caption{Instantaneous mechanical advantage for the $2\times 2$ lattice quantum lever where only the level $\varepsilon_1$ is driven (protocol 1). Here the temperature $T=0.0001$, chemical potential $\mu=0$, coupling $w=1$, and $\varepsilon_2=-10$, $\varepsilon_3=-2$, $\varepsilon_4=-10$. }
    \label{fig_2by2_e1driven_mechadv}
\end{figure}

In order to explore the possible universality of this bound on $\eta$, we have also computed the mechanical advantage of a quantum lever base on a four-level quantum system, a $2\times 2$ lattice.
The Hamiltonian %
is given by Eq.\ \eqref{eq_closedsys_ham_gen} with %
\begin{equation}
 H_{S}(t)=\varepsilon_{1}(t)d_1^{\dagger}d_1 \,, 
 \end{equation}
\begin{equation} 
H_{\bar{S}}=\sum_{i=2}^{4}\varepsilon_{i}d^{\dagger}_id_i+ \sum_{i=2}^{3} w (d^{\dagger}_id_{i+1} + \mbox{h.c.}) \,,
\end{equation}
and 
\begin{equation}
H_{S\bar{S}}=\sum_{i=\{2,4\}} w (d^{\dagger}_1d_{i} + \mbox{h.c.}) \,. 
\end{equation}

Figure \ref{fig_2by2_e1driven_mechadv} shows the mechanical advantage of a quantum lever based on this $2\times 2$ lattice, where again level 1 is driven and the work on subspace 1 is computed.  Similar to the 2-level quantum lever, a maximum mechanical advantage of $\eta \approx 2$ is obtained for this 4-level system.  An analytical argument based on a perturbative analysis of an $N$-level system in the low-temperature limit is presented in Appendix \ref{app_lever_bound}, and confirms the bound $\eta \leq 2$.  We believe the finding of a 100\% mechanical advantage is significant, but it is an open question whether other configurations could lead to yet higher values of the work amplification.

\section{Conclusions}\label{sec_conclusions}

In this work, we have analyzed the thermodynamics of a subsystem of a larger (but finite) quantum system driven quasi-statically by external forces. The system as a whole is treated in the grand canonical ensemble, exchanging particles and energy with a macroscopic reservoir that is itself in thermal equilibrium.  The specific example analyzed involves a system of independent fermions, but the framework presented for partitioning thermodynamic quantities is equally applicable to bosonic systems.

Our analysis of the thermodynamics of a quantum subsystem %
is based on the Hilbert-space partitioning scheme that has recently been established for the thermodynamics of open quantum systems \cite{kumarWorkSumRule2024a,webbHowPartitionQuantum2024a}.  For the present case of a subsystem of a finite quantum system, the expressions derived for the internal energy \eqref{eqn_part_energy_def}, entropy \eqref{eqn_part_entropy_def}, occupancy \eqref{eqn_part_partnum_def}, and grand potential \eqref{eqn_part_grandpot_def} have the advantage of simplicity of interpretation, wherein the contribution of each instantaneous energy eigenstate of the system is weighted by the probability to find a particle in that eigenstate residing within the subsystem.  Quantum work, on the other hand, eludes such a simple interpretation.  Nonetheless, the expression for nonlocal work in closed systems \eqref{eqn_part_omegacurr_prob_rel} offers additional insights compared to the more opaque Green's function expressions \cite{kumarWorkSumRule2024a} for open systems, evincing both free energy flow and generation terms.

Our partitioning of the thermodynamics reveals %
the nonlocal character of the total external quantum work done. This is encapsulated in a sum rule for quantum work [Eq.\ \ref{eqn_closedsys_worksumrule}], which equates the total external work done to the sum of the partitioned work done on all the subsystems. The nonlocal aspect of quantum work is most evident when the external drive acts locally on a subsystem and yet thermodynamic work is performed, in a quantum nonlocal manner, on the complementary subsystem as well.  

We then investigated how nonlocal work can be used to generate local amplification of the total external work done. We termed such a system a quantum lever which amplifies not only the local force but also the local work performed.  By simulating this quantum lever action of two- and four-level quantum systems, we demonstrated that by driving a given subsystem of a bipartite quantum system one can produce as much as a 100\% increase in the work done on that subsystem by tuning the parameters so as to perform negative nonlocal work on the complementary subspace.

Our partitioning scheme thus sheds light not only on thermodynamics of closed quantum systems but also on quantum control. Since we only considered quasi-static driving in this paper, it will be interesting to generalize the scheme presented here to arbitrary driving in future work.

\acknowledgements

We thank Caleb Webb for helpful discussions as well as for collaboration on related works \cite{kumarWorkSumRule2024a,webbHowPartitionQuantum2024a}. We also thank Carter Eckel, Ferdinand Evers, and Yiheng Xu for insights developed during the preliminary stages of this work.

\bibliography{%
cqs.bbl
}

\clearpage

\onecolumngrid

\appendix

\section{Full Hamiltonian in the limit of quasi-static driving}\label{app_full_ham_qstat_drive}
The dynamics of the universe including the reservoir is described by the Hamiltonian 
\begin{equation}
    H_{tot}(t)= H(t)+ H_R+ H_{S\bar{S}-R}
\end{equation}
where $H(t)$ is defined in Eq.\ \eqref{eq_closedsys_ham_gen}, $H_R$ models the macroscopic reservoir that is maintained at temperature $T$ and chemical potential $\mu$ and $H_{S\bar{S}-R}$ models the weak coupling between the reservoir and the composite system ($S+\bar{S}$) of interest. 

The equation of motion of any observable $O(t)$ acting on the system Hilbert space is given by 
\begin{equation}
    \frac{d}{dt}\langle O(t)\rangle=\langle\dot{O}(t)\rangle+\frac{1}{i\hbar}\langle [H_{tot}(t),O(t)]\rangle ,
\end{equation}
and because $[O(t), H_R]=0$, this may be written as
\begin{equation}
    \frac{d}{dt}\langle O(t)\rangle=\langle\dot{O}(t)\rangle+\frac{1}{i\hbar}\langle [H(t),O(t)]\rangle 
    + \frac{1}{i\hbar}\langle [H_{S\bar{S}-R}(t),O(t)]\rangle.
\end{equation}
The limit of quasi-static driving holds when 
\begin{eqnarray}
    |H|\gg|H_{S\bar{S}-R}|\gg \frac{\hbar|\dot{H}|}{|H|}.
\end{eqnarray}
In this limit, $H_{S\bar{S}-R}$ does not alter the eigenstates or eigenenergies of $H(t)$, but merely mediates the incoherent exchange of particles and energy between the system and reservoir.  Moreover, when the driving is sufficiently slow, this simply maintains the system in equilibrium with the reservoir at all times, leading to the following result valid to 
 first-order in the quasi-static limit:
\begin{equation}
\frac{1}{i\hbar}\langle [H_{S\bar{S}-R}(t),O(t)]\rangle^{(1)}
  =\sum_{\nu}\dot{f}(\epsilon_\nu(t))\langle \nu(t)|O(t)|\nu(t) \rangle.
\end{equation}

\section{Connection between Hilbert-space partition and subspace Probability-weighted expressions }\label{app_hilbertspace_part_derv}

To show how taking the quantum-statistical average of the Hilbert-space partitioned observables [Eq.\ \eqref{eqn_hilb_sp_partition_def}] leads to subspace probability-weighted expressions [Eqs.\ \eqref{eqn_part_energy_def}, \eqref{eqn_part_entropy_def}, \eqref{eqn_part_partnum_def}, and \eqref{eqn_part_grandpot_def}], we start by first considering a dynamical observable $O$ that is compatible with the Hamiltonian $H(t)$. The quantum statistical average of the partitioned observable $O|_\gamma$ can be written as
\begin{equation}
    \langle O|_\gamma(t)\rangle ^{(0)}=\sum_{\nu}\langle\nu(t)|o|_\gamma|\nu(t)\rangle f(\epsilon_{\nu}(t))\,,
\end{equation}
which by definition [Eq.\ \eqref{eqn_hilb_sp_partition_def}] can be written as
\begin{equation}
    \langle O|_\gamma(t)\rangle ^{(0)}=\sum_{\nu}\langle\nu(t)|\frac{1}{2}\{o,\pi_\gamma\}|\nu(t)\rangle f(\epsilon_{\nu}(t))\,.
\end{equation}
Due to the compatibility of $O$ and $H$, this becomes
\begin{equation}
    \langle O|_\gamma(t)\rangle ^{(0)}=\sum_{\nu}\langle\nu(t)|\pi_\gamma|\nu(t)\rangle f(\epsilon_{\nu}(t))o(\epsilon_{\nu}(t))\,,
\end{equation}
where from Eq.\ \eqref{eqn_prob_nu_gamma_def} we immediately identify 
\begin{equation}
    \langle O|_\gamma(t)\rangle ^{(0)}=\sum_{\nu}P_{\nu}(\gamma,t) f(\epsilon_{\nu}(t))o(\epsilon_{\nu}(t))\,.
\end{equation}
This establishes the connection between Hilbert-space partitioning and subspace probability-weighted expressions for 
quantum dynamical quantities that are compatible with the Hamiltonian, such as the particle number and (trivially) the energy. Statistical observables like entropy can also be shown to obey the same partitioning scheme (for details see Ref.\ \cite{webbHowPartitionQuantum2024a} and Sec.\ VII of Supplemental Material for Ref.\ \cite{kumarWorkSumRule2024a}). 

An alternative route to obtaining expectation values of local observables is in terms of the local spectrum of a subsystem \cite{shastryTemperatureVoltageMeasurement2016,shastryThirdLawThermodynamics2019,kumarWorkSumRule2024a}. The local spectrum is related to the local density of states (LDOS) measured by various scanning probe techniques \cite{%
binnigScanningTunnelingMicroscopy1983,binnigAtomicForceMicroscope1986,kalininScanningProbeMicroscopy2007,chenIntroductionScanningTunneling2021,muraltScanningTunnelingPotentiometry1986,williamsScanningThermalProfiler1986,shiThermalProbingEnergy2009,kimQuantitativeMeasurementScanning2011,yuHighresolutionSpatialMapping2011,kimUltraHighVacuumScanning2012a,mengesQuantitativeThermometryNanoscale2012,Lee2013,mecklenburgNanoscaleTemperatureMapping2015,mengesTemperatureMappingOperating2016,staffordLocalTemperatureInteracting2016,shastryTemperatureVoltageMeasurement2016,Shastry2020_STTh}, and  is defined for subsystem $\gamma$ as 
\begin{equation}
    g^{(0)}_\gamma (\epsilon,t) = \int_{x\in\gamma} dx\, \langle x|\delta(\epsilon-h(t))|x\rangle\,. 
\end{equation}
The LDOS is simply the spatial resolution \cite{shastryThirdLawThermodynamics2019,kumarWorkSumRule2024a} of the familiar total density of states (the number of states available in the composite system ($S+\bar{S}$) per unit energy). The LDOS is related to the 
probability of finding the particle in state $\nu$ within subsystem $\gamma$ via the spectral theorem, using Eq.\ \eqref{eqn_prob_nu_gamma_def}, as
\begin{equation}
    g^{(0)}_\gamma (\epsilon,t) = \sum_\nu P_{\nu}(\gamma,t) \delta(\epsilon-\epsilon_\nu(t))\,.
\end{equation}

\section{Derivation of the expression for partitioned power %
[Eq.\ \eqref{eqn_part_extwork_prob_rel}]}
\label{app_closedsys_extpower_prob_rel_derv}

The quantum statistical average of the partitioned power operator %
in the quasi-static limit is given by
\begin{eqnarray}
    \langle \dot{H}|_\gamma(t)\rangle&=&\sum_{\nu}f(\epsilon_{\nu})\langle\nu|\dot{h}|_{\gamma}|\nu\rangle \nonumber \\
    &=& \frac{1}{2}\sum_{\nu}f(\epsilon_{\nu})\langle\nu|\dot{h}\pi_{\gamma}+\pi_{\gamma}\dot{h}|\nu\rangle \nonumber \\
    &=& \frac{1}{2}\sum_{\mu,\nu}f(\epsilon_{\nu}) \left[\langle\nu|\dot{h}|\mu\rangle\langle\mu|\pi_{\gamma}|\nu\rangle+\langle\nu|\pi_{\gamma}|\mu\rangle\langle\mu|\dot{h}|\nu\rangle\right] \nonumber \\
    &=& \sum_{\mu,\nu}\frac{f(\epsilon_{\nu})+f(\epsilon_{\mu})}{2}\langle\nu|\dot{h}|\mu\rangle\langle\mu|\pi_{\gamma}|\nu\rangle\,,
\end{eqnarray}
from which Eq.\ \eqref{eqn_part_extwork_prob_rel} follows immediately, using for the diagonal term %
$\langle \nu|\dot{h}(t)|\nu\rangle=\dot{\epsilon}_\nu(t)$.

\section{Contrasting nonlocal work with energy transfer mediated by the coupling $H_{S\bar{S}}$}
\label{app_nonlocalwork_vs_usualegytransfer}

A natural question that arises is how the phenomenon of nonlocal work differs from the usual flow of energy between the two systems mediated by the coupling Hamiltonian $H_{S\bar{S}}$. To highlight the contrast between the two, we explicitly compute the rate of change of the internal energy of the subsystem $\bar{S}$.  
It follows from the analysis of Appendix \ref{app_full_ham_qstat_drive} that we may write the equation of motion for the expectation value of the Hamiltonian partitioned on subsystem $\bar{S}$ as 
\begin{equation}
    \frac{d}{dt}\langle H|_{\bar{S}}\rangle=\frac{1}{2}\langle \dot{H}_{S\bar{S}}\rangle +\frac{1}{i\hbar}\langle[H,H|_{\bar{S}}]\rangle+\sum_{\nu}\dot{f}(\epsilon_\nu(t))\langle \nu(t)|H|_{\bar{S}}|\nu(t) \rangle,
\end{equation}
where the energy transfer mediated by $H_{S\bar{S}}$ is given by the second term on the rhs. To first order in quasi-static driving, wherein the expectation values in the above equation are evaulated using $\rho^{(0)}(t)$,
this term vanishes. This can be seen by noting that with $\rho^{(0)}(t)=e^{-\beta(H(t)-\mu N -\Omega)}$, we may write 
\begin{eqnarray}
  \langle[H,H|_{\bar{S}}]\rangle^{(1)}&=&\mathrm{Tr}\{\rho^{(0)}[H,H|_{\bar{S}}]\}\nonumber\\&
  =&\mathrm{Tr}\{\rho^{(0)}H H|_{\bar{S}}- \rho^{(0)}H|_{\bar{S}} H\}\,.
\end{eqnarray}
Clearly, $[\rho^{(0)}(t),H(t)]=0$, so that we may write
\begin{equation}
    \langle[H,H|_{\bar{S}}]\rangle^{(1)}=\mathrm{Tr}\{H \rho^{(0)} H|_{\bar{S}}- \rho^{(0)}H|_{\bar{S}} H\}\,,
\end{equation}
which, owing to the cyclicity of the trace, vanishes. 

It is then clear that in the quasi-static limit the ``ordinary" flow of energy into $\bar{S}$ mediated by $H_{S\bar{S}}$ is zero, whereas the nonlocal work done is non-zero as discussed extensively in Sec. \ref{sec_wsr}.

\section{Quasi-statically driven two-level system } \label{app_tls_exact_exprs}
 For quasi-static driving of the two-level system all the thermodynamic quantities can be written explicitly. We illustrate this for the total  external power delivered and the partitioned internal energy (of subsystem 1). For the total external power we have from Eq. \eqref{eqn_qstat_extpower_spec_rel_b}

 \begin{equation}
  \dot{W}_{\text{ext}}  = f_+\dot{\epsilon}_{+} +f_-\dot{\epsilon}_{-},
 \end{equation}
where the eigenvalues are given by
\begin{equation}
  \epsilon_{\pm}=\frac{\epsilon_1+\epsilon_2}{2}\pm\sqrt{\Big(\frac{\Delta}{2}\Big)^2+w^2}  
\end{equation}
with $\Delta=\epsilon_1-\epsilon_2$, and $f_\pm \equiv f(\epsilon_\pm)$ denotes the Fermi-Dirac function evaluated at the eigenvalues.

For the internal energy of subsystem 1 we can write from Eq.\ \eqref{eqn_part_energy_def} (suppressing the time variable)
\begin{equation}
    U_1^{(0)}=\sum_{\nu={\pm}}P_{\nu}(1)f_\nu\epsilon_{\nu},
\end{equation}
where the partitioned probability may be calculated as
\begin{equation}
   P_{\pm}(1)= |\langle 1|\pm\rangle|^2=1-\frac{4w^2}{(\Delta\pm\sqrt{\Delta^2+4w^2})^2+4w^2},
\end{equation}
where $|\pm\rangle$ are the eigenvectors of the the two-level system and $|1\rangle$ denotes one of the two vectors in the dot/spatial basis. The expression for the rate of change of the partitioned internal energy $\dot{U}_1^{(1)}(t)$, while cumbersome, can be obtained from the above expression by simply using the product rule. Furthermore, all other partitioned quantities for the two-level system can be obtained in closed form in an analogous manner. 

\section{Partitioned thermodynamics for the two-level system driven according to protocol 1}\label{app_part_thermo_prot1}

\begin{figure}
    \centering
    \includegraphics[width=0.45\textwidth,height=19cm]{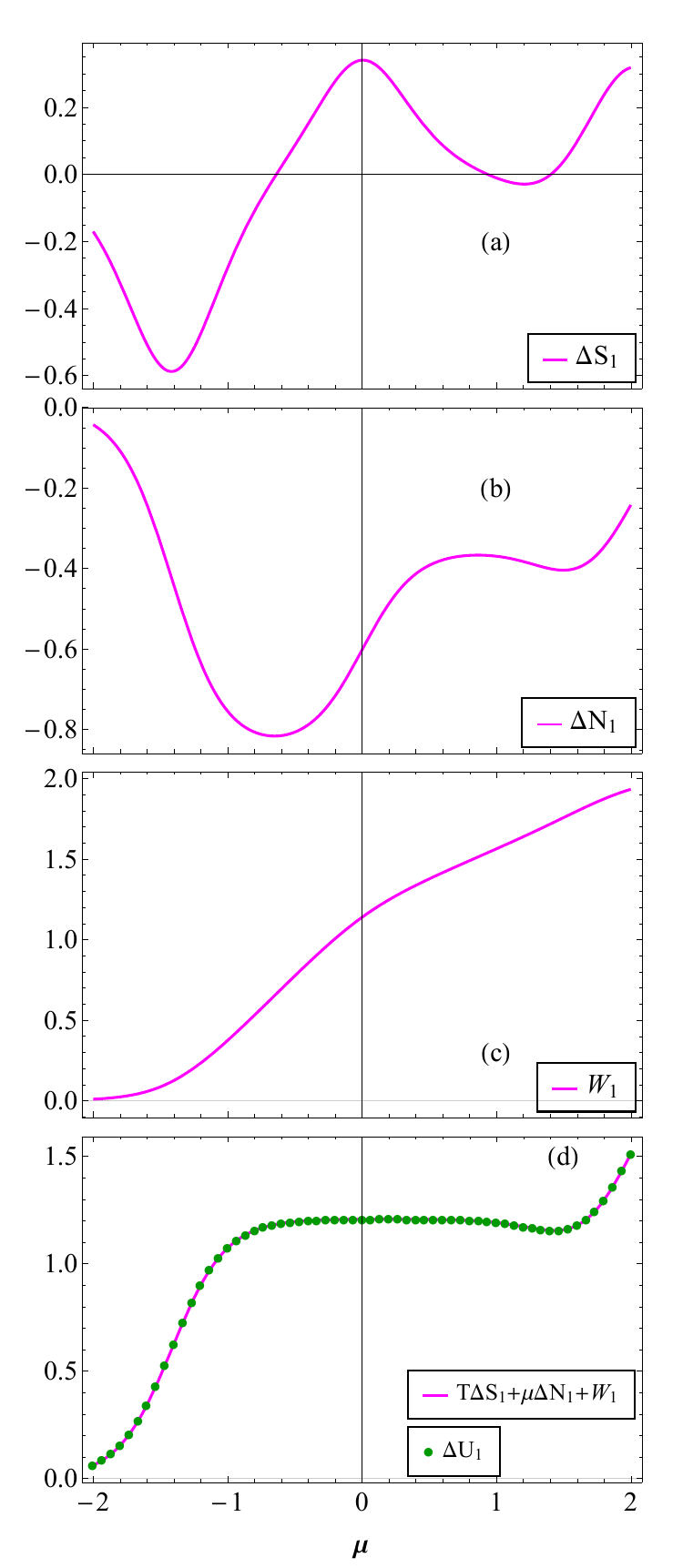}
    \captionsetup{justification=raggedright, singlelinecheck=false}
    \caption{The integrated terms (a) entropy $\Delta S_1$, (b) occupancy $\Delta N_1$, and (c) work $W_1$
in the two-level model under driving protocol 1, as functions of the chemical potential of the reservoir. (d) Verification of the equality of the LHS and RHS of Eq.\ \eqref{eqn_1st_Law_gamma}.  Here the reservoir temperature $T=0.2$, and  the left onsite energy is varied as $\varepsilon_1$: $-1\rightarrow 1$, while keeping the other site energy and coupling fixed at $\varepsilon_2=1$ and $w=1$, repsectively.}
\label{fig_tls_E1driven_firstlaw_grid}
\end{figure}

Fig.\ \ref{fig_tls_E1driven_firstlaw_grid} presents the partitioned thermodynamics for the two-level system driven according to protocol 1.

\section{Further discussion of Fig.\ \ref{fig_tls_E1Vdriven_firstlaw_grid}} \label{app_tls_thermo_fig_discuss}

The negative peak in the change in entropy of subsystem 1, $\Delta S_1$, around $\mu=-1.4$ in Fig.\ \ref{fig_tls_E1Vdriven_firstlaw_grid}(a) arises due to the depopulation of the bonding orbital of the 2-level system during the protocol, while the positive peak in $\Delta S_1$ around $\mu=+1.0$ arises due to the increase in entropy of both the bonding and antibonding orbitals at the end of the protocol \footnote{From Eq.\ \eqref{eqn_part_entropy_def}, at $\mu=-\sqrt{2}$ the entropy at the beginning of the protocol is dominated by that of the half-filled bonding orbital that is localized mainly on subsystem 1, and the entropy at the end of protocol tends toward zero as the system is depopulated. At $\mu=+1.0$, the entropy at the beginning of the protocol is nearly zero as the bonding orbital is nearly full and the antibonding orbital is nearly empty, while the entropy is maximal at the end of the protocol as the bonding and antibonding orbitals are nearly delocalized over the two subsystems and nearly half full.}.
The behavior of $\Delta N_1$ in Fig.\ \ref{fig_tls_E1Vdriven_firstlaw_grid}(b) 
can be understood in terms of the occupancy of the system before and after the driving protocol:  For $\mu<-1.4$, the bonding orbital localized mainly on site 1 is occupied before the protocol and empty afterwards, while for $\mu>1.1$, the bonding orbital is filled throughout the protocol while the antibonding orbital, initially localized mainly on site 2, is occupied after the protocol and delocalized over both sites. The partitioned work $W_1$ in Fig.\ \ref{fig_tls_E1Vdriven_firstlaw_grid}(c) is an increasing function of $\mu$, tending to zero as $\mu\rightarrow -2$ as the system becomes empty, and saturating at +2 for $\mu>1.4$ when the system is doubly occupied throughout the protocol.

\section{Lever bound derivations}\label{app_lever_bound}

\subsection{Two-level system}
It follows from Eq.\ \eqref{eqn_part_grandpot_prob_rel} that the instantaneous mechanical advantage [Eq.\ \eqref{eqn_mech_adv_def}] for the two-level system can be written as 
\begin{equation}
\eta=\frac{\dot{W}_{1}}{\dot{W}_{\text{ext}}}=\frac{\sum_{\nu=\{+,-\}}[\dot{P}_{\nu}(1,t)\omega_{\nu}+P_{\nu}(1,t)f_\nu\dot{\epsilon}_{\nu}]} {\sum_{\nu=\{+,-\}}f_\nu\dot{\epsilon}_{\nu}}
\end{equation}
Motivated by the optimal parameters found in a numerical exploration of the mechanical advantage of the 2-level quantum lever, it is useful to do a perturbative analysis in the low-temperature limit
to estimate a bound on $\eta$. 
To lowest non-vanishing order in $w/\Delta$, we can write the eigenvalues, eigenvectors and probabilities for the two-level system as 
\begin{equation}
    \epsilon_{+}\simeq \epsilon_{1}+\frac{w^2}{\Delta}\,,
\end{equation}
\begin{equation}
    \epsilon_{-}\simeq  \epsilon_{2}-\frac{w^2}{\Delta}\,,
\end{equation}
\begin{equation}
    |+\rangle \simeq  |1\rangle + \frac{w}{\Delta}|2\rangle\,,
\end{equation}
\begin{equation}
    |-\rangle \simeq |2\rangle - \frac{w}{\Delta}|1\rangle\,,
\end{equation}
\begin{equation}
   P_{+}(1)= |\langle 1|+\rangle|^2\simeq 1-\Big(\frac{w}{\Delta}\Big)^2\,,
\end{equation}
and 
\begin{equation}\label{eqn_P_minus_approx}
   P_{-}(1)= |\langle 1|-\rangle|^2\simeq \Big(\frac{w}{\Delta}\Big)^2\,.
\end{equation}
These results allow us to write  
\begin{equation}
    \eta \simeq \frac{\dot{\epsilon}_1[P_+^2(1)f_+ + P_-^2(1)f_-] + \dot{P}_-(1)(\omega_- - \omega_+)}{\dot{\epsilon}_1[P_+(1)f_+ + P_-(1)f_-]}\,,
\end{equation}
where we have used that 
\begin{equation}
    \dot{\epsilon}_+ \simeq \dot{\epsilon}_1 - \Big(\frac{w}{\Delta}\Big)^2\dot{\epsilon}_1 = P_+(1)\dot{\epsilon}_1\,,
\end{equation}
and 
\begin{equation}
    \dot{\epsilon}_- \simeq \Big(\frac{w}{\Delta}\Big)^2\dot{\epsilon}_1 = P_-(1)\dot{\epsilon}_1\,.
\end{equation}
Now as $T\rightarrow0$, we have $f_+ \rightarrow 0$ and $\omega_+ \rightarrow 0$, so that we may write in this limit  
\begin{equation}
    \eta \simeq \Big(\frac{w}{\Delta}\Big)^2+\frac{\dot{P}_-(1)\omega_-}{\dot{\epsilon}_1 P_-(1)f_-}\,.
\end{equation}
Dropping the negligible $\Big(\frac{w}{\Delta}\Big)^2$ term and rewriting  $\frac{\dot{P}_-(1)}{P_-(1)}=\frac{d}{dt}(\ln P_-(1))$ we get 
\begin{equation}
  \eta \simeq \Big(\frac{\omega_-}{f_-}\Big)\frac{d}{dt}(\ln P_-(1))\frac{1}{\dot{\epsilon}_1}\,.
\end{equation}
Upon using Eq.\ \eqref{eqn_P_minus_approx}, we can write $\frac{d}{dt}(\ln P_-(1))=-2\frac{\dot{\epsilon}_1}{\Delta}$, so that we have
\begin{equation}
  \eta \simeq -2\Big(\frac{\omega_-}{f_-}\Big)\frac{1}{\Delta}\,.
\end{equation}
Now, as $T\rightarrow0$, we also have $f_-\rightarrow1$, which in turn implies that  $\frac{\omega_-}{f_-}\rightarrow\epsilon_- -\mu$, the above expression becomes %
\begin{equation}
    \eta \simeq 2\Big(\frac{\mu-\epsilon_-}{\Delta}\Big)\,,
\end{equation}
which is approximately 
\begin{equation}
     \eta \simeq 2\Big(\frac{\mu-\epsilon_2}{\epsilon_1-\epsilon_2}\Big) \leq 2
\end{equation}
since $\epsilon_1<\mu<\epsilon_2$, %
thus proving the upper limit on the mechanical advantage claimed in the main text.

\subsection{Multi-level system}

The result of the previous subsection holds for quasi-statically driven multi-level systems as well. For such systems the instantaneous mechanical advantage may be written as 
\begin{equation}\label{eqn_mechadv_mls_def}
\eta=\frac{\dot{W}_{S}}{\dot{W}_{\text{ext}}}=\frac{\sum_{\mu=\{\nu,\bar{\nu}\}}[\dot{P}_{\mu}(S,t)\omega_{\mu}+P_{\mu}(S,t)f_\mu\dot{\epsilon}_{\mu}]} {\sum_{\mu=\{\nu,\bar{\nu}\}}f_\mu\dot{\epsilon}_{\mu}}\,,
\end{equation}
where $\{\nu\}$ represents the manifold of states primarily localized in subsystem $S$, and $\{\bar{\nu}\}$ the manifold of states primarily localized in the complementary subsystem $\bar{S}$.
We will choose the system local driving to be of the form 
\begin{equation}
    H_S(t)=H_{S}^{0}+\Delta(t)\Pi_S\,,
\end{equation}
and the average coupling between two subsytems as 
\begin{equation}
  \overline{ \langle \nu| H_{S\bar{S}}|\bar{\nu}\rangle}^2=w^2 \,,
\end{equation}
such that $w\ll\Delta(t)$. 

To lowest non-vanishing order, the eigenstates maybe written 
\begin{equation}
    |\nu\rangle\simeq|\nu^{(0)}\rangle + \sum_{\nu\neq\bar{\nu}}\frac{\langle\bar{\nu}^{(0)
    }|H_{S\bar{S}}|\nu^{(0)}\rangle}{\varepsilon_\nu^{(0)}-\varepsilon_{\bar{\nu}}^{(0)}}|\bar{\nu}^{(0)}\rangle \,,  
\end{equation}
where we have omitted time labels to lighten notation. Using the above and the fact that the projector onto the subspace $S$ is given by $\Pi_S=\sum_{\nu\in S}|\nu^{(0)}\rangle\langle\nu^{(0)}|$, we may evaluate the probability $P_{\bar{\nu}}(S):=\langle\bar{\nu}|\Pi_S|\bar{\nu}\rangle$ as
\begin{equation}\label{eq_P_nubar_expr}
  P_{\bar{\nu}}(S,t)\simeq\left(\frac{w}{\Delta}\right)^2 \mathrm{dim}\{\bar{S}\}\,,
\end{equation}
so that from normalization at the same order we may write
\begin{equation}
    P_{\nu}(S,t)\simeq \mathrm{dim}\{S\} - \left(\frac{w}{\Delta}\right)^2 \mathrm{dim}\{\bar{S}\}\,.
\end{equation}
Now as $T\rightarrow0$, we have $f_\nu \rightarrow 0$ and $\omega_\nu \rightarrow 0$, so that in this limit, Eq.\ \eqref{eqn_mechadv_mls_def} becomes
\begin{equation}
    \eta\simeq\frac{\sum_{\bar{\nu}\in \bar{S}}\dot{P}_{\bar{\nu}}(S,t)\omega_{\bar{\nu}}}{\sum_{\bar{\nu}\in \bar{S}}f_{\bar{\nu}}\dot{\epsilon}_{\bar{\nu}}}+\frac{\sum_{\bar{\nu}\in \bar{S}}P_{\bar{\nu}}(S,t)f_{\bar{\nu}}\dot{\epsilon}_{\bar{\nu}}}{\sum_{\bar{\nu}\in \bar{S}}f_{\bar{\nu}}\dot{\epsilon}_{\bar{\nu}}}\,.
\end{equation}
The second term in the above equation clearly evaluates to simply $P_{\bar{\nu}}(S)\simeq\left(\frac{w}{\Delta}\right)^2 \mathrm{dim}\{\bar{S}\}$, which we can ignore, whereas for the first term we can write, using $\dot{\epsilon}_{\bar{\nu}}=\dot{\Delta}P_{\bar{\nu}}(S)$ and noting from Eq.\ \eqref{eq_P_nubar_expr} that $P_{\bar{\nu}}(S)$ is approximately independent of $\bar{\nu}$, that 
\begin{equation}
\eta\simeq\left(\frac{\dot{P}_{\bar{\nu}}(S,t)}{\dot{\Delta}P_{\bar{\nu}}(S,t)}\right)\left(\frac{\sum_{\bar{\nu}\in \bar{S}}\omega_{\bar{\nu}}}{\sum_{\bar{\nu}\in \bar{S}}f_{\bar{\nu}}}\right)\,,
\end{equation}
which evaluates to 
\begin{equation}
    \eta \simeq \left(\frac{-2}{\Delta}\right)\left(\frac{\mathrm{dim}\{\bar{S}\}\overline{\omega_{\bar{\nu}}}}{\mathrm{dim}\{\bar{S}\}\overline{f_{\bar{\nu}}}}\right)\,.
\end{equation}
In the low-temperature limit we note that,  $f_{\bar{\nu}}\rightarrow 1$ (which implies $\omega_{\bar{\nu}}\rightarrow \epsilon_{\bar{\nu}}-\mu$) and the above becomes
\begin{equation}
    \eta \simeq \left(\frac{-2}{\Delta}\right)\left(\overline{\epsilon_{\bar{\nu}}-\mu}\right)\,,
\end{equation}
which for $\epsilon_\nu<\mu<\epsilon_{\bar{\nu}}$ implies, similar to the two-level case in the previous subsection, that $\eta\leq2$. 

\newpage
\pagestyle{empty}

\end{document}